# Ray-Tracing Calibration from Channel Sounding Measurements in a Millimeter-Wave Industrial Scenario

Grégory Gougeon[1], Frédéric Munoz[2,3], Yoann Corre[1], Raffaele D'Errico[2,3]
[1] SIRADEL, Saint-Grégoire France, ggougeon@siradel.com
[2] CEA-LETI, Grenoble, France, frederic.munoz@cea.fr
[3] University of Grenoble-Alpes, Grenoble, France

*Abstract*—New-generation communication and sensing systems are gaining strong interest in the context of Industry 4.0 e.g., related to mapping techniques, environmental sensing, automation or hyper-vision. The radio propagation in confined, cluttered and heavily metalized factory environments is a critical challenge; thus an evaluation by accurate propagation channel models is necessary. Site-specific channel emulation can be obtained from Ray-tracing (RT); but RT validation for factory environments is still an on-going work. For this purpose, a measurement campaign was performed in a machine room with many metallic objects and machines, using a mmWave channel sounder. Wideband channel responses were collected and compared to RT simulations. The RT prediction tool was calibrated to minimize the error observed on some large scale statistics, thus reaching a very good agreement between the simulation and the measurement. Average error in received power, delay spread and azimuth spread is below 1.5 dB, 5 ns and 2° respectively.

## I. INTRODUCTION

3GPP interest in channel modelling for industrial environment started only recently with release 15.1 [1] at the end of 2019. A new scenario of interest, namely « Indoor Factory (InF) », was introduced, with five sub-scenarios depending on the clutter density and terminals height [1]. Clutters represent various kinds of industry furniture (e.g. machines, assembly lines, storage shelves). The different sub-scenarios have distinct models for large-scale parameter models (e.g. Line-of-Sight (LoS) probability and path loss), while the small-scale parameters modelling (i.e. multi-path components) only distinguishes between LoS and Non-LoS (NLoS). Most of the 3GPP indoor factory results are summarized in [2], with contributions coming partly from the measurement campaign described in [3]. In brief, the 3GPP model has been calibrated based on measurements at 3.5 GHz and 28 GHz. In [4], the measurements in an industrial environment at 3.5 GHz show that the percentage of total channel power attributed to non-specular components can reach 64% in NLoS, and is 4% to 13% greater than in an office environment due to the cluttered and metallic nature of the environment. Besides, it is observed that strong coverage may be obtained even in NLoS areas. The average small-scale parameters extracted from measurements in [5], [6] are in tight agreement with the values predicted by the 3GPP model, as well as the mean cross-polarization ratio (XPR), evaluated between 12.8 dB and 18.0 dB. However, some of these measurements show that the path-loss models strongly depend on the specific scenario. The analysis of the propagation channel in [7] at 28 GHz and 60 GHz demonstrated scenario-dependent small-scale parameters and led to the conclusion that ray-based tools can be useful for reliable predictions.

Contrary to the stochastic and empirical channel models such as proposed by 3GPP, the ray-tracing (RT) techniques require the environment to be precisely modeled, and the antennas to be positioned at relevant locations. This approach permits to accurately reproduce channel characteristics consistent with the physical environment i.e. the specificities of a particular factory will be considered. Both sub-7GHz and millimeter-wave (mmWave) frequencies are usually well supported. However, before application of the RT tool in any industrial scenario, proper calibration and qualification are needed [8]. This can be obtained by comparison to channel sounder measurements, from which several channel properties can be extracted. The multi-dimensional aspect of the calibration problem is a challenge [8], [9], [10], but this approach offers full characterization.

In this paper we exploit a mmWave channel sounder measurement campaign realized in a machine room to calibrate an indoor RT model. The measurement setup is described in section II. The main RT principles are given in section III. The model assessment and calibration are detailed in section IV. Conclusions are drawn in section V.

## II. CHANNEL SOUNDER MEASUREMENT CAMPAIGN

### A. Channel sounder description

Propagation channel measurements were performed in a machine room using a vector network analyzer. The propagation channel was probed from 26 to 30 GHz, with a sampling rate of 5 MHz (i.e. 801 frequency points). The power transmitted to the transmit antenna port (Tx), was set to 10 dBm. On the receiver side (Rx) a vertically polarized monopole antenna was used while on the Tx side a virtual antenna array was measured. This array consists in a vertically polarized monopole antenna mounted on an automatic positioner to scan a 5x5x5 virtual cubic array in the x, y and z axes.

Since the center frequency is 28 GHz, a grid spacing of 5 mm, corresponding to half a wavelength, was considered. The antennas were placed at height of 1.52 m above the ground. The measurement instruments were controlled by an external computer. Fig. 1 shows the measurement setup.

### B. Environment and measurement scenario

Fig. 2 shows the layout of the measurements in the machine room which dimensions are 35 m long, 16 m wide and 3.5 m high. Different propagation conditions were considered: direct view (LoS), and obstructed (OLoS / NLoS), covering a range of relative distance Tx-Rx from 1 to 10 m. In all scenarios, the antenna array is placed at a fixed position while the Rx antenna changes position. In the LoS scenarios, the measurements were performed along the machine room aisle where the roof is constructed with metal pipes and metal bars (see Fig. 3a); 10 equally spaced points were measured there for Tx-Rx distances varying from 1 m to 10 m with a step size of 1 m. For the OLoS (obstructed) scenarios, 7 points were probed for Tx-Rx distances ranging from 5 m to 11 m with a 1 m step. The Rx antenna was moving in a metal corridor behind a metal cabinet higher than the measurement antennas where the roof of this part of the room is entirely metal (see Fig. 3b). In the NLoS scenarios, large ventilation shafts and soundproof pads covered with metal sheets blocked the direct LoS path; the environment includes closed electrical cabinets with metal doors (see Fig. 3c). Subsequently, 3 points with Tx-Rx distances of 5 m to 7 m were probed. In the OLoS and NLoS measurements performed with virtual antenna arrays, a low noise amplifier with a gain of 20 dB is connected to the Rx antenna.

### C. Measurement post-processing description

From the bi-static measurements, the Multi Path Components (MPCs) extraction was performed by the so-called Space Alternating Generalized Expectation Maximization (SAGE) high-resolution algorithm. Only MPCs with significant powers of 3 dB above background noise were considered. In this context, we assume the noise behaves as a normal distribution with zero mean. It is expected to be independent and consistently distributed across each acquisition. Two stopping criteria were considered: 1) the maximum number of MPCs to be extracted is set to 100 and; 2) when more than 95% of the measured channel power is reached [11], [12].

### III. CHANNEL MODELING

#### A. Ray-tracing channel model description

The Volcano Flex [10] ray-tracing is a time-efficient propagation model capable to predict a deterministic path-loss in any small-scale urban or indoor scenario, and to provide channel properties and 3D multi-path trajectories. The ray-tracing relies on the precise description of the environment given in the 3D digital map. Rays are launched from the Tx position towards the objects that compose the building(s) to produce one or several successive interactions such as reflections, diffractions, transmissions and scattering, then reaching possibly the Rx location. Rays undergo direction changes, attenuation and depolarization computed from the Geometrical Optics and Uniform Theory of Diffraction, considering dielectric characteristics provided through ITU recommendations [13].

The validity of Volcano Flex in industrial scenarios is verified by a comparison to the mmWave radio channel measurements described in section II. A calibration permits to minimize the gap between simulations and measurements.

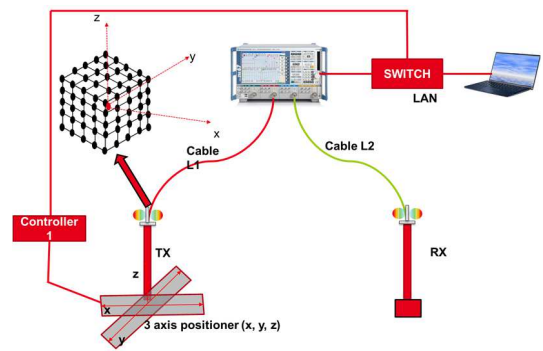

Fig. 1. Measurement setup with virtual antenna array 5x5x5.

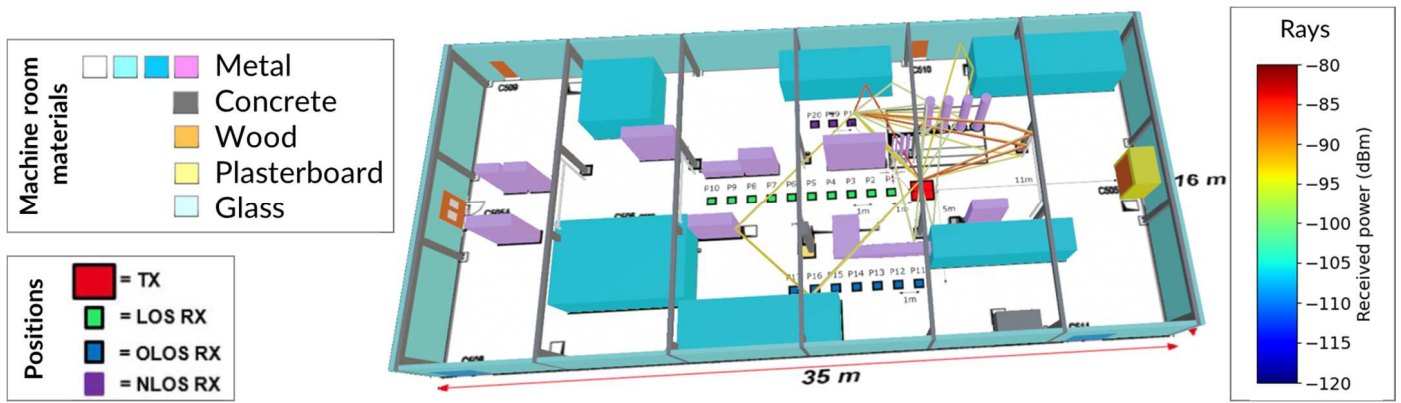

Fig. 2. Measurement points; and main predicted rays for NLoS Rx 18.

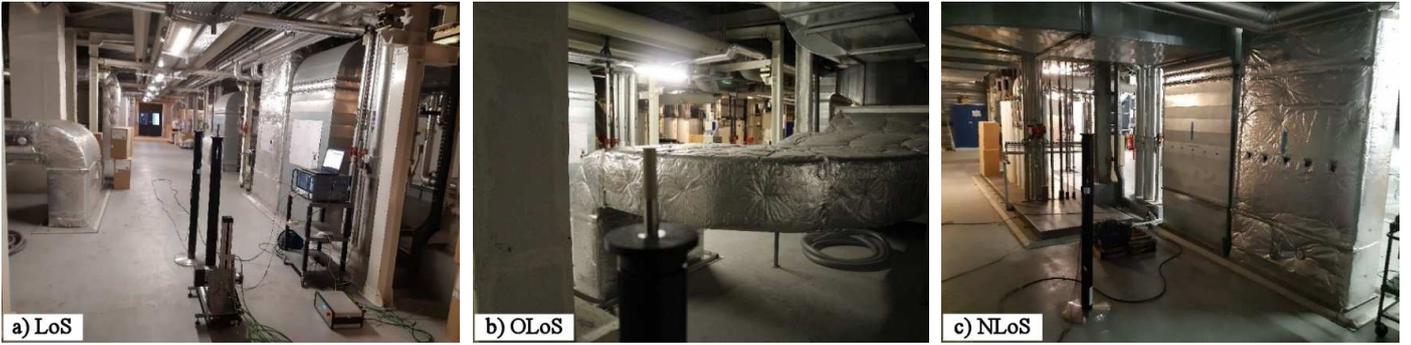
Fig. 3. Pictures of the measurements.

## B. Digital mapping of the environment and antennas

Accurate digital map is essential for successful comparison between measurements and simulations, in particular at the high considered frequency. A 3D twin of the machine room of size 16×35 m and height 3.5 m was constructed from the 2D architect floor plan and some photos taken during the measurement campaign. As shown in Fig. 2, the contour walls, ceiling, machines, pipes and some pillars are made of metal. The ground and other pillars and beams are made of concrete. There is no window, and doors are made of wood. Thus the considered room is a confined highly reflecting environment.

The Tx and 20 Rx were positioned in the digital map (Fig. 2) thanks to the distances reported on the 2D floor plan during the measurements, and with *a posteriori* adjustments:
- For LoS Rx: the Tx-Rx distance is adjusted according to the propagation delay of the measured direct path,
- All Rx's of each sub-scenario (LoS, OLoS or NLoS) are aligned along a straight trajectory,
- Two successive Rx's are separated by approx. 1 m,
- Rx11 position is slightly moved such that a pillar obstructs the direct path, in accordance with the low-power first-arriving path observed in the measurements.

## C. Simulation parameters

Both Tx and Rx are equipped with a vertically-polarized half-wave monopole antenna. They are positioned at 1.52 m above the ground. The simulation frequency is set to 28 GHz. The Volcano Flex model is configured such that maximum 2 reflections and 2 diffractions can be generated along a propagation path. Diffuse scattering is enabled for the metallic machines, pipes and pillars.

## IV. SIMULATION AND CALIBRATION RESULTS

### A. Correction of the digital map

Correction of the digital environment was a necessary stage at the beginning of the study, since physical details may have a significant impact on the observed propagation. Two examples are given below to illustrate this preliminary work.

Following the methodology of [10], we first realized a mapping between the predicted rays and the measured contributions, based on the available channel characteristics: received power; propagation delay; and emitting azimuth. Then we identified rays that are missing in the prediction or, on the contrary, rays not found in the measurement. This discrepancy can be fixed by precisely checking photos of the real environment, and bringing corrections in the digital map. We added one long pipe below the ceiling, passing over the OLoS positions. This pipe blocks the unexpected long-delay rays, which were reflecting on both a wall and the ceiling before reaching the LoS Rx positions.

Fig. 4 (left) shows the measured Power Angular Delay Profile obtained at NLoS Rx 18, where the main emitting propagation directions can be easily identified. Besides, Fig. 4 (right) plots some of the predicted ray trajectories leaving the Tx position (red square); one can see strong paths propagating between the pipes (in purple). The measured emitting azimuths confirm the existence of such paths in the 330-360° range, but with much weaker power. The prediction is too optimistic when rays propagate between two close pipes. For this reason, a virtual envelope was created in the digital map to add 20 dB/m attenuation to the paths propagating in the pipes vicinity. This brings the expected local correction.

### B. Simulations from the initial model

The accuracy of the initial RT model (before calibration) is assessed by comparing measured and predicted Large Scale Parameters (LSP): total power, mean delay, delay spread, mean azimuth, and azimuth spread. The LSP values are evaluated from the specular contributions only, i.e. the contributions extracted by SAGE on the measurements side, and the reflected and diffracted rays on the predictions side. Results are given in Fig. 5 (see the red dotted line for the initial prediction outcomes). A good match is obtained for the received power, with an error standard deviation of 1.64 dB and 1.81 dB for respectively the LoS positions and OLoS+NLoS positions. Regarding the delay spread, it is of 2.56 ns and 4.07 ns respectively, while for the azimuth spread it is of 9.43° and 14.62 respectively.

The percentage of diffused power is another metric that was compared. The initial prediction of the diffuse component is found to be strongly underestimated, leading to an average of 32% missing power at OLoS or NLoS positions. One explanation comes from the SAGE algorithm. Indeed, the number of extracted specular contributions is limited to 100; and the whole remaining power is considered as diffused. This limit was reached for most measured Rx positions; thus a part of the measured diffuse power may not have been properly categorized. However, the main reason for the underestimated diffused power is more likely the lack of predicted diffuse contributions at long ranges, as illustrated later in this section.

Detailed analysis permits to have a more physical understanding of the differences between measurements and simulations. For example, Fig. 6 shows the Power Distance Profile (PDP) obtained at the LoS position Rx 5. The specular components extracted by the SAGE algorithm from the measurement are displayed in blue, together with the predicted specular components in orange. The received power is displayed against the propagation distance instead of the delay in order to easily connect the paths with the physical environment. The strong contributions at 40-45 meter propagation distance have been identified as rays propagating along the machine room aisle, then reflecting back to Rx from the left boundary wall. These contributions appear in both the measurement and prediction results. Note that the predicted propagation distance is shorter compared to the measurement, which might be due to discrepancy between the digital map and reality, but has not been corrected here. The interesting point is that the predicted power is higher than observed in the measurement, which means that the simulation of the reflection coefficients is too optimistic. Besides, the PDP shows that the power of the predicted diffuse components rapidly decreases at larger delays, which results in many missing long-range contributions observed in all Rx positions in the simulation.

*C. Model calibration*

The calibration consists in the optimal adjustment of two offset values (expressed in dB), which are applied respectively on each reflection coefficient and each diffraction coefficient; the objective is to globally minimize the LSP RMSE errors. The coefficients of the diffuse interactions are corrected accordingly. A consistent power balance between the reflected and diffused powers is preserved. The correction is such that the sum of those two powers is kept constant when changing the reflection offset. The optimal reflection offset is found to be -3 dB, while the best diffraction offset is -2 dB, improving the prediction of most the LSP's (see the green line of Fig. 5).

The statistics on LoS LSP's are given in Tab.I ; the ones for OLoS+NLoS are given in Tab. II . The prediction of the total specular received power was quite good already before the calibration; however, the RMSE is reduced further for all visibility conditions. The calibration permits to reduce the error standard deviation of the LoS delay spread from 2.56 ns to 1.75 ns. It increases from 4.07 ns to 4.34 ns for the OLoS+NLoS positions, however the mean error (1.47 ns to -0.69 ns) and correlation (0.52 to 0.67) are significantly improved. Regarding the spatial dimension of the channel, the error standard deviation of the LoS azimuth spread is reduced from 9.43° to 6.17°, while for the OLoS+NLoS positions it is reduced from 14.62° to 14.04°. It is worth noting the model is able to reproduce the large OLoS+NLoS azimuth spread: 40.79° predicted vs. 42.59° measured. Finally, the prediction of the diffused power percentage is improved but remains underestimated, with now 22% of the total received power missing at OLoS or NLoS positions.

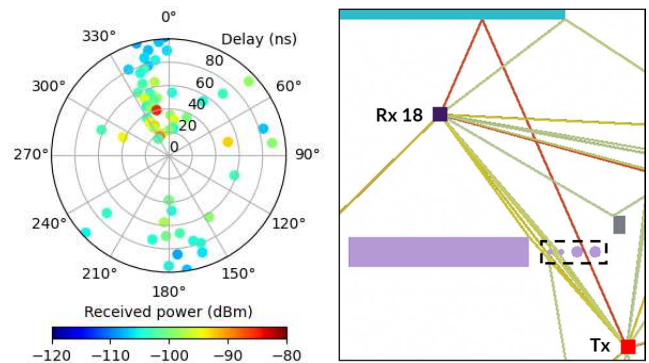

Fig. 4. Results at NLoS Rx 18. Left: Measured Power Emitting Angular Delay Profile. Right: Zoom on predicted specular rays from Fig. 2. The same color legend is used for the ray power levels in the two plots.

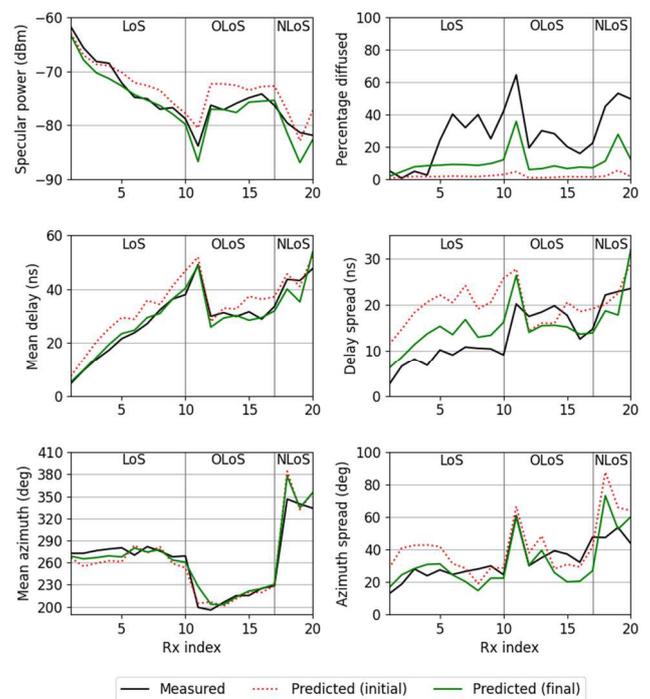

Fig. 5. Measured and predicted large scale parameters.

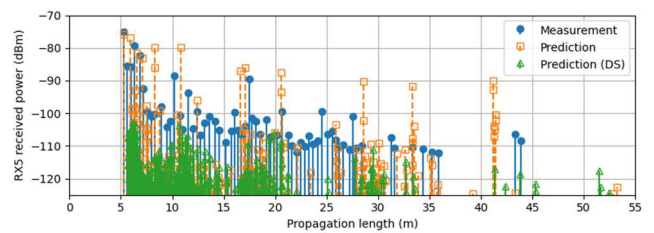

Fig. 6. Comparison of measurement and prediction at LoS Rx 5.

TABLE I. RESULTS FOR LOS LARGE SCALE PARAMETERS

| LS Statistic | Measured | Predicted (initial) | Predicted (final) |
|---|---|---|---|
| Total specular power (dBm) | -71.89 | -70.99 | -72.96 |
| Mean error (dB) | | 0.90 | -1.07 |
| Standard deviation (dB) | | 1.64 | 1.10 |
| RMSE (dB) | | 1.87 | 1.53 |
| Correlation | | 0.97 | 0.98 |
| Percentage diffused (%) | 21.55 | 1.61 | 7.86 |
| Mean error (%) | | -19.94 | -13.69 |
| Standard deviation (%) | | 15.65 | 14.27 |
| RMSE (%) | | 25.35 | 19.78 |
| Correlation | | 0.64 | 0.70 |
| Specular delay spread (ns) | 8.35 | 19.64 | 12.73 |
| Mean error (ns) | | 11.29 | 4.38 |
| Standard deviation (ns) | | 2.56 | 1.75 |
| RMSE (ns) | | 11.57 | 4.74 |
| Correlation | | 0.79 | 0.84 |
| Specular azimuth spread (°) | 24.31 | 33.17 | 23.39 |
| Mean error (°) | | 8.86 | -0.92 |
| Standard deviation (°) | | 9.43 | 6.17 |
| RMSE (°) | | 12.94 | 6.23 |
| Correlation | | -0.05 | 0.27 |

TABLE II. RESULTS FOR OLOS+NLOS LARGE SCALE PARAMETERS

| LS Statistic | Measured | Predicted (initial) | Predicted (final) |
|---|---|---|---|
| Total specular power (dBm) | -78.16 | -75.45 | -79.64 |
| Mean error (dB) | | 2.72 | -1.48 |
| Standard deviation (dB) | | 1.81 | 1.69 |
| RMSE (dB) | | 3.26 | 2.25 |
| Correlation | | 0.87 | 0.94 |
| Percentage diffused (%) | 34.71 | 2.00 | 12.79 |
| Mean error (%) | | -32.71 | -21.92 |
| Standard deviation (%) | | 14.91 | 9.03 |
| RMSE (%) | | 35.95 | 23.71 |
| Correlation | | 0.77 | 0.87 |
| Specular delay spread (ns) | 18.87 | 20.35 | 18.18 |
| Mean error (ns) | | 1.47 | -0.69 |
| Standard deviation (ns) | | 4.07 | 4.34 |
| RMSE (ns) | | 4.33 | 4.39 |
| Correlation | | 0.52 | 0.67 |
| Specular azimuth spread (°) | 42.59 | 49.96 | 40.79 |
| Mean error (°) | | 7.36 | -1.80 |
| Standard deviation (°) | | 14.62 | 14.04 |
| RMSE (°) | | 16.37 | 14.16 |
| Correlation | | 0.67 | 0.65 |

## V. CONCLUSION

The presented work consists in the evaluation and calibration of a RT propagation model in a complex and specific industrial environment. For this study, channel sounding data was measured around 28 GHz in a machine room for LoS, OLoS and NLoS situations. The specular components extracted by SAGE were compared to the Volcano Flex RT prediction. The prediction errors on various LSP parameters were minimized in two steps: 1) corrections in the digital map; and 2) calibration of the interactions strength. Finally, the mean error in the specular received power, delay spread and azimuth spread is below 1.5 dB, 5 ns and 2° respectively for all LoS, OLoS and NLoS positions. The corresponding LoS error standard deviations are 1.10 dB, 1.75 ns and 6.17°. The OLoS+NloS error standard deviations are 1.69 dB, 4.34 ns and 14.04°. The RT performance is found to be globally good compared to the measurements.

Prediction of the diffuse component is more complicated, and remains significantly underestimated for long propagation ranges, even after calibration.

The ray-tracing tool will now be used to test, assess and demonstrate the performance of wireless sensing and mapping solutions deployed in various factory environments. In this context, the limited accuracy for prediction of long-delay diffuse contributions might be of negligible impact. But the calibration work must be extended to include the validation and adjustment of the cross-polarization ratios; this on-going task relies on a multi-polar measurement campaign that is carried out in same factory environment as presented in this article.

ACKNOWLEDGMENT

This work has been funded by the French National Agency for Research (ANR) under the S²LAM project ANR-21-CE25-0017-01.

REFERENCES

[1] 3GPP, "Study on channel model for frequencies from 0.5 to 100 GHz", TR 38.901 V15.1.0 (2019-09).
[2] T. Jiang et al., "3GPP Standardized 5G Channel Model for IIoT Scenarios: A Survey," in IEEE Internet of Things Journal, 2020.
[3] J. Narrainen and R. D'Errico, "Large Scale Channel Parameters in Industrial Environment," EuCAP 2019, Krakow, Poland.
[4] Tanghe et al., "Experimental analysis of dense multipath components in an industrial environment", IEEE TAP 2014.
[5] S. Jaeckel et al., "Industrial Indoor Measurements from 2-6 GHz for the 3GPP-NR and QuaDRiGa Channel Model," IEEE VTC, Honolulu, USA, 2019.
[6] M. Schmieder et al., "Measurement and Characterization of an Indoor Industrial Environment at 3.7 and 28 GHz," EuCAP 2020, Copenhagen, Denmark.
[7] D. Solomitckii, A. Orsino, S. Andreev, Y. Koucheryavy and M. Valkama, "Characterization of mmWave Channel Properties at 28 and 60 GHz in Factory Automation Deployments," IEEE WCNC, Barcelona, Spain, 2018.
[8] G.S. Bhatia, Y. Corre, M. Di Renzo, "Tuning of Ray-Based Channel Model for 5G Indoor Industrial Scenarios," IEEE MeditCom, Dubrovnik, Croatia, 2023.
[9] M. Z. Aslam, et al., "Analysis of 60-GHz In-street Backhaul Channel Measurements and LiDAR Ray-based Simulations," EuCAP 2020, Copenhagen, Denmark.
[10] R. Charbonnier et al., "Calibration of Ray-Tracing With Diffuse Scattering Against 28-GHz Directional Urban Channel Measurements," in IEEE Transactions on Vehicular Technology, Dec. 2020.
[11] A. Mudonhi, G. Makhoul, M. Lotti, R. D'Errico, and C. Oestges, "Mmwave massive mimo channel sounding in industrial iot scenarios," in Joint EuCNC/6G Summit, 2022.
[12] F. Munoz, G. Makhoul, D. P. Gaillot, C. Oestges and R. D'Errico, "Space-Time Dense Multipath Components Modeling at mmWaves in Indoor Industrial Environments," EuCAP 2023, Florence, Italy.
[13] ITU, Recommendation ITU-R P.2040-1, "Effects of building materials and structures on radiowave propagation above about 100 MHz", July 2015.